\documentstyle[prb,aps,epsfig]{revtex}
\begin{document}
\baselineskip=12pt

\draft
\flushbottom
\twocolumn[\hsize\textwidth\columnwidth\hsize
\csname@twocolumnfalse\endcsname
\title {
Double Exchange Models:
Self Consistent Renormalisation}

\author{Sanjeev Kumar and Pinaki Majumdar}

\address{ Harish-Chandra  Research Institute,\\
 Chhatnag Road, Jhusi, Allahabad 211 019, India }

\date{May 22, 2005}

\maketitle
\tightenlines
\widetext
\advance\leftskip by 57pt
\advance\rightskip by 57pt
\begin{abstract}

We propose a scheme for constructing classical spin Hamiltonians from Hunds 
coupled spin-fermion models in the  limit $J_H/t \rightarrow \infty$. The 
strong coupling between fermions and the core spins requires  self-consistent 
calculation of the effective exchange in the model, either in the presence 
of inhomogeneities or with changing temperature.  In this paper we establish
the formalism and discuss  
results mainly on the ``clean'' double exchange model,
with self consistently renormalised 
couplings, and compare  our results with exact simulations.
Our method allows access to system sizes much beyond the reach of exact 
simulations, and  we can study transport and optical properties of the model 
without artificial broadening. 
The method discussed here forms the foundation of our papers
Phys. Rev. Lett. {\bf 91}, 246602 (2003), and 
Phys. Rev. Lett. {\bf 92}, 126602 (2004).

\

\

\end{abstract}

]

\narrowtext
\tightenlines

\section{Introduction}

The double exchange (DE) model was introduced by Zener \cite{zener}
in 1951  to
motivate ferromagnetism in the perovskite manganites. In contrast 
to `Heisenberg like' coupling between localised spins, the
effective interaction in `double exchange' arises from optimisation
of carrier kinetic energy in the spin background. 
The intimate correlation between spin configuration and 
electron motion had,
till recently,
restricted the
study of the DE model to mostly qualitative analysis or 
mean field theory. The original proposal of  Zener was followed
up \cite{and-has}
 by Anderson and Hasegawa, who clarified the physics of the
coupled spin-fermion system in a two site model,
and  de Gennes \cite{degennes}
who presented a thermodynamic
calculation and  a phase diagram (incorporating 
antiferromagnetic superexchange). He produced
the first estimate of transition temperature $(T_c)$ in the
model. The thermodynamic transition within double exchange was
also studied \cite{kub-ohata}
by Kubo and Ohata.
This short list essentially exhausts activity on the double
exchange problem prior to the `manganite renaissance'.

The discovery of colossal magnetoresistance (CMR) and
a variety of magnetic phases  in the
manganites \cite{mang-rev}
led to renewed  interest in the DE  
model. In addition, the availability of powerful
analytical and numerical tools, {\it e.g}, dynamical
mean field theory (DMFT) and Monte Carlo methods 
provided impetus for studying the DE 
model in detail.
In real systems the  double exchange 
interaction is supplemented by \cite{millis-tokura}
antiferromagnetic (AF) superexchange,
electron-phonon interactions,
 and disorder, and some of these models
have been studied within various approximations.
The primary limitation of current methods, as we discuss in detail later,
is their inability to access transport properties taking spatial
fluctuations and disorder effects fully into account. In this context
our method, of constructing an approximate but {\it explicit} classical
spin Hamiltonian, allows a breakthrough.
In the present paper our detailed results are  on the simplest
case, of the clean DE model.
In earlier short publications we have presented results on the 
disordered double exchange model \cite{sk-pm-dde}, and on
magnetic phase competetion \cite{sk-pm-nano1}.

Let us define the general model to which our method is
applicable.
$H= H_{el} + H_{AF}$, with 
\begin{eqnarray}
H_{el}~ &=&
\sum_{\langle ij \rangle, \sigma} t_{ij} 
c^{\dagger}_{i \sigma} c^{~}_{j \sigma}
+  \sum_{i } (\epsilon_i - \mu) n_{i}  
- J_H\sum_i {\bf S}_i {\bf .} {\vec \sigma}_i  \cr
H_{AF} &=& J_S\sum_{\langle ij \rangle} {\bf S}_i 
{\bf .} {\bf S}_j 
\end{eqnarray}

The $t_{ij} =-t$ 
are nearest neighbour hopping, on a square or cubic lattice as
relevant. 
$\epsilon_i$ is the 
 on site potential,  uniformly distributed between $\pm \Delta/2$, say,
and $J_S$ is an antiferromagnetic superexchange between the core spins.
$J_H$ is the `Hunds' coupling, and we will work in the limit
$J_H/t \rightarrow \infty$.
The parameters in the problem are $\Delta/t$, $J_S/t$,
and the carrier density $n$ (or chemical potential $\mu$).
We assume 
a classical core spin, setting
$\vert {\bf S}_i \vert =1$, and absorb the magnitude of the spin
in $J_S$.
All our energy scales, frequency $(\omega)$ and temperature $(T)$,
 etc, will be measured in units
of $t$.

For  $J_H/t \rightarrow \infty$ the fermion spin at a site
is constrained to be parallel 
to the core spin,
gaining energy $-J_H/2$, while the `antiparallel'
orientation is pushed to $+J_H/2$. 
Since the hopping term $t_{ij}$ itself is spin conserving,
the motion of the low energy, locally parallel spin, fermions 
is now controlled by nearest neighbour spin orientation.
The strong magnetic coupling $(J_H)$ generates an
effective single band `spinless fermion' problem \cite{dex-proj}, 
with core spin orientation dependent hopping amplitudes. 
We will discuss the hopping term 
further on, for the moment let us denote the renormalised (spin 
orientation dependent) 
hopping amplitude  as $\tilde t$, 
indicative of double-exchange physics.

\vspace{.2cm}

The ${\tilde t}-\Delta-J_S$ problem has a variety of ground states.
$(i)$~In the absence of $J_S$, both the `clean' and 
the disordered DE  model 
has a {\it ferromagnetic ground state}, at all electron density,
 with $T_c$ reducing with increase
in $\Delta$. 
$(ii)$~The non disordered problem, with $J_S$, leads to
a variety of phases \cite{dex-js-2d,dex-js-3d}
competing with ferromagnetism.
These are {\it ferromagnetic and A, C, 
G type AF phases, etc}. There could also be more exotic `flux',
`skyrmion' or `island' phases in some parts of parameter space.
The  boundaries between these phases are often 
first order so there are regimes of 
{\it macroscopic phase coexistence}.
The specific set of possible AF phases depends on $J_S$.
$(iii)$~
{\it Weak } disorder in the ${\tilde t} - J_S$ problem
\cite{sk-pm-nano1,dag-clust}
converts the regions of macroscopic phase separation into {\it mesoscopic
phase coexistence of FM and AF clusters.} 
$(iv)$~
For some density and $\Delta-J_S$ combination, the 
ground state could be  a 
{\it spin glass}. 

Although the phases above can be motivated easily, the electrical
character of the ground state, or the temperature dependence of 
magnetic and transport properties, or the response 
to an applied magnetic field, are still not well understood. 
A comprehensive
understanding of these effects
within the relatively simple model in Eqn.1
would be the first step in approaching 
the even richer variety of phases in the manganites, where the 
lattice degrees of freedom are also active.
This calls for a new technique, handling spatial and thermal 
fluctuations, the formation of 
clusters, and the effect of   
electron localisation. 
We propose and extensively benchmark such a real space 
technique in this paper. To appreciate the need for a new 
method
let us quickly review the current approaches to the Hamiltonian 
above.

\subsection{Theoretical approaches}

The approaches can be broadly classified into three categories.
These are: 
$(i)$~Exact variational calculations \cite{var-t0-maek}
at $T=0$, and 
generalisation \cite{var-tf-varm,var-tf-golo,var-tf-guin}  
to $T \neq 0$ via approximate mean field techniques. Let us
call these methods variational mean field (VMF), for 
convenience.
$(ii)$~Dynamical mean field theory (DMFT) based calculations
\cite{dmft-furuk,dmft-kubo}
which map on the lattice model 
to an effective single site problem
in a temporally fluctuating medium. Apart from a formal limit
$d \rightarrow \infty$, where $d$ is the number of spatial 
dimensions, there are no further approximations in the theory.
$(iii)$~Real space, finite size, Monte Carlo (MC) simulations 
\cite{ed-mc-dag,ed-mc-cal-br,ed-mc-furu,furu-new,nucl-phys}
of the coupled `spin-fermion' problem, treating the core spin as
classical.

We can set a few indicators in terms of which the strength and
weakness of various approximations can be judged. These are,
tentatively: 
\begin{enumerate}
\item
The ability to access ground state properties.
\item
Ability to handle fluctuations, 
and accuracy
of $T_c$ estimate.
\item
The ability to access response functions, {\it e.g},
transport and 
optical properties.
\item
Treatment of disorder effects: 
Anderson localisation  and cluster coexistence.
\item
Ability to handle Hubbard interactions, and
quantum effects in spins and  phonons.
\item
Computational cost and  finite size effects. 
\end{enumerate}

\subsubsection{Variational calculations}

The variational calculations attempt a minimisation of the energy
of the (clean) system, at $T=0$, with respect to a variety of ordered
spin configurations. The optimal
configuration $\{ {\bf S}_i \}_{min}$ for
specified $J_S$, $\mu$, etc,
is accepted as the magnetic ground state. 
The energy calculations are relatively straightforward, since the
electron motion is in a periodic background.
The
method has been used to map out the ground state phase diagram 
of DE model with AF superexchange in two and three 
dimension \cite{dex-js-2d,dex-js-3d}.
The approach, however, can only be approximately implemented
at finite temperature
\cite{var-tf-varm,var-tf-golo,var-tf-guin}.  
One has to calculate a spin distribution
instead of just targeting the ground state,
and estimating the energy of an electron system in a spin
disordered background is non trivial.
Due to the mean field character of VMF, fluctuation effects are
lost and transition temperatures are 
somewhat overestimated.
The method is focused on thermodynamic properties so
there is no discussion of  transport, etc,
within this scheme (with one exception \cite{var-tf-varm}).
Disorder effects have been included,
approximately \cite{var-tf-varm}, 
in some of these calculations.
Variational methods can provide indication of phase coexistence 
\cite{dex-js-2d,dex-js-3d}
at $T=0$, or,
approximately, at finite temperature
\cite{var-tf-golo,var-tf-guin}, but cluster coexistence
in a disordered system is beyond its reach. The method has
not been generalised to include quantum many body effects. 
Finite size effects in this approach 
 are small and the method is relatively
easy to implement.

\subsubsection{Dynamical mean field theory}

The single site nature of the DMFT approximation becomes
 exact in the limit of 
`high dimensions'.
DMFT can access both ground state and finite temperature
properties, but the effective single site approximation
cannot capture spatial fluctuations, or a non trivial
paramagnetic phase. The `mean field' character 
leads to an overestimate of $T_c$,
and also the inability to differentiate between two and
three dimensional systems.
Being a Greens function based theory DMFT can 
readily access
response functions. However, effects like Anderson 
localisation or cluster coexistence, which require spatial
information,
 cannot be accessed\cite{dmft-andloc}. The method can handle many body,
Hubbard like, interactions and quantum dynamics in all the 
variables involved, although such calculations are quite
difficult.
DMFT is defined directly in the thermodynamic limit, so
there are no finite size effects. The calculations are 
relatively easy, when quantum many body effects are not
involved, and have been a major tool in exploring 
phenomena in the manganites.

The  limitations of DMFT become
apparent as we consider the more complicated phases that
can arise in our
model. For instance in the  strong disorder problem \cite{sk-pm-dde},
when there is a 
possibility of electron localisation, the DMFT approach
cannot access the insulating  phase \cite{dmft-andloc}. 
Neither can it access the spatially inhomogeneous nature of
freezing, and the persistence of strong spin correlations
above the bulk $T_c$.
Similarly, in the problem of competing double exchange and
superexchange, in the presence of weak disorder, the system
breaks up into interspersed `ferro-metallic' and `AF-insulating'
regions \cite{sk-pm-nano1,dag-clust}. A complicated variant of this coexistence effect 
has been extensively 
studied in manganite experiments \cite{ph-coex-expts}. 
The `single site' nature of DMFT
cannot access cluster coexistence, except possibly in an averaged
sense. The transport and metal-insulator transitions that
can occur in this situation also remain inaccessible.
So, there are important {\it qualitative effects}
beyond the reach of DMFT,
in systems where spatial inhomogeneity is important.

\subsubsection{Monte Carlo}

The finite size real space approach uses the Metropolis
algorithm to generate equilibrium configurations of the
spins at a given temperature. Monte Carlo calculations
on classical systems with 
short range interactions involve a cost ${\cal O}(zN)$ for
a system update, with $z$ being the coordination number on
the lattice and $N$ the system size. In the spin-fermion
problem, however, the `cost' of a spin update at a site
has to be computed from the fermion free energy. 
If one uses direct diagonalisation of the
Hamiltonian to accomplish this, the cost {\it per site}
is ${\cal O}(N^3)$, the cost for a `system update'
is a prohibitive $N^4$.
All this is after ignoring quantum many body effects. 
Current MC approaches have not been generalised to
handle Hubbard like interactions.

Despite the severe computational cost, this method,
which we will call ED$-$MC (exact diagonalisation based MC),
has been successfully used to clarify several aspects of 
manganite physics, and DE models
in general. System sizes accessible are $\sim 100$
at most (recent algorithms \cite{furu-new} have enhanced this somewhat),
with $50-60$ being more typical. 
This method can provide an outline of the finite 
temperature magnetic 
phase diagram, reveal major
spectral features, and even yield the 
basic signatures of cluster coexistence.
However, as is obvious 
from the accessible $N$, the finite size gaps are
much too large for any reasonable estimate of d.c 
transport properties, and the small linear dimension
available, in two or three spatial dimension,
allows only
a preliminary glimpse of  coexistence physics. 
The size limitation apart, the method is {\it exact}
and comprehensive, with none of the problems of standard quantum
Monte Carlo (QMC).
An extension of this approach 
to larger system sizes would allow
exploration of several unresolved issues in manganite physics.
Apart from the ED based MC, `hybrid MC' results have been
reported \cite{dex-js-3d,nucl-phys} for the various phases of
double exchange competing with superexchange antiferromagnetism.
No transport results, however, have yet been presented within this
framework.

Our method,  described  in the next section,
is developed in this spirit.
It is  a {\it real space Monte Carlo approach} with the
key advantage that it avoids the iterative $N^3$ 
diagonalisation step. We extract an effective Hamiltonian
for the core spins from the coupled spin-fermion problem,
through a self-consistent scheme. We can work at
arbitary temperature, handle strong disorder, and
have better control on `cluster physics' and transport
properties due to our significantly larger system size,
$N \sim 10^3$.

In the next section we describe our approximation and its
computational implementation in detail. Following that we
describe our results on 
the  `clean' DE 
model in two and three dimension.
 We will discuss 
results on thermodynamics, spectral features, resistivity and optical
conductivity, in most of these cases, and 
compare with exact simulation
results. We will also highlight systematically 
the size effects
in transport and optical properties. 

\section{Method}

\subsection{The $J_H/t \rightarrow \infty $ limit}

We have already written down our basic Hamiltonian in Eqn.1.
The transformation and projection described in the next couple
of paragraphs 
is well known, but we repeat them here for completeness.

Working at large $J_H/t$ it is useful to `diagonalise'
the $J_H {\bf S}_i.{\vec \sigma}_i$ term first.The electron
spin operator is ${\vec \sigma}_i = \sum_{\alpha \beta}
c^{\dagger}_{i \alpha} {\vec \sigma}_{\alpha \beta} c_{i \beta}$,
where the 
$  {\sigma}^{\mu}_{\alpha \beta} $ are the Pauli matrices,  
and this  $2 \times 2$ problem has eigenvalues $\pm J_H/2$. The
eigenfunctions are linear combinations of the standard
`up' and `down'  $z$ quantised 
fermion states at the site:
$ \gamma^{\dagger}_{i\mu} = \sum_{\alpha} A^i_{\mu \alpha}
c^{\dagger}_{i \alpha}$.
 The lower energy state, $\gamma^{\dagger}_{il}$,
a linear combination of the form 
$ A^i_{11} c^{\dagger}_{i \uparrow} + A^i_{12} c^{\dagger}_{i \downarrow}$,
is at energy $-J_H/2$ and 
has fermion spin parallel to the core spin ${\bf S}_i$. The 
orthogonal linear combination, $\gamma^{\dagger}_{iu}$,
 has fermion spin anti-parallel to the
core spin and is at energy $+ J_H/2$.
The amplitudes $A^i_{\mu\alpha}$ are standard \cite{dex-proj}.

In the $\gamma$ basis, the Hunds coupling term becomes $-(J_H/2)
( \gamma^{\dagger}_{il}\gamma_{il} - \gamma^{\dagger}_{iu}\gamma_{iu})$
at all sites. The intersite hopping term, however, picks up a 
non trivial dependence on nearest neighbour spin orientation,
$t_{ij} c^{\dagger}_{i\sigma}c_{i\sigma} \rightarrow 
\sum_{\alpha \beta} 
t_{ij} g^{\alpha \beta}_{ij} \gamma^{\dagger}_{i \alpha} 
\gamma_{j \beta}$
where $\alpha,\beta$ refer to the $u, l$ indices.
$g^{\alpha \beta}_{ij}$ arises from the product of the
two transformations at site $i$ and site $j$, and we will
describe its specific form later.
 Since the canonical
transformation is local, the density operator
$\sum_{\sigma} c^{\dagger}_{i \sigma} c_{i \sigma} 
\rightarrow  (\gamma^{\dagger}_{i l} \gamma_{il} + \gamma^{\dagger}_{iu}
\gamma_{iu})$.

At finite $J_H/t$ this is just a transformation from the `lab
frame' to a local axis and the `up' and `down' spin fermions get
mapped to $(l,u)$, but we still have to solve a mixed `two orbital problem'.
However, if $J_H/t \rightarrow \infty$ then all
the `anti-parallel' 
$\gamma^{\dagger}_{iu} \vert 0 \rangle$ states get projected 
out and we can work solely in the subspace of states created by
$\gamma^{\dagger}_{il}$. In this space, the Hamiltonian assumes
a simpler form:

\begin{eqnarray}
H_{el} & =&  
-t\sum_{\langle ij \rangle} 
(~g_{ij}  \gamma^{\dagger}_i  \gamma_j + 
h.c~) + 
\sum_i (\epsilon_i - \mu) n_i \cr
&=& -t\sum_{\langle ij \rangle} f_{ij}
(~e^{i \Phi_{ij}}  \gamma^{\dagger}_i  \gamma_j + 
h.c~) + 
\sum_i (\epsilon_i - \mu) n_i
\end{eqnarray}
where we have dropped the superfluous $ll$ label in $g_{ij}$,
and absorbed $-J_H/2$ in the chemical potential.
The hopping amplitude $g_{ij} = f_{ij} e^{i\Phi_{ij}}$ 
between locally aligned states,
can be written in terms of the polar angle $(\theta_i)$ and
azimuthal angle $(\phi_i)$ of the spin ${\bf S}_i$ 
as,
$  cos{\theta_i \over 2} cos{\theta_j \over 2}$ 
$+
sin{\theta_i \over 2} sin{\theta_j \over 2}
e^{-i~(\phi_i - \phi_j)}$.
It is easily checked that 
the `magnitude'  of the overlap is
$f_{ij} = \sqrt{( 1 + {\bf S}_i.{\bf S}_j)/2 }$,
while the phase is specified by 
$tan{\Phi_{ij}} = Im(g_{ij})/Re(g_{ij})$.

This problem can be viewed as a quadratic `spinless fermion'
problem with core spin dependent hopping amplitudes.
The fermions move in the 
background of quenched disorder $\epsilon_i$ and
`annealed disorder' in the $\{ {\bf S}_i \}$, where the second
brackets indicate the full spin configuration. 
To exploit
the nominally `non interacting' structure  of the fermion part
we need to know the relevant spin configurations, $\{ {\bf S}_i \}$,
or, more generally, the 
distribution
$P \{ {\bf S}_i \}$, 
controlling  the probability of occurence of a  spin 
configuration. 

\subsection{Effective Hamiltonian for spins}

The~partition function of the system is  $Z = \int {\cal D}{\bf S}_i
Tr e^{-\beta H}$. 
To extract 
$P \{ {\bf S}_i \}$
note that for a system with only spin degrees of freedom, $Z$ will
have the form
$\int {\cal D}{\bf S}_i
e^{-\beta H \{ {\bf S} \}}$. Comparing this with the 
partition function of the spin-fermion problem we can use 
$$\int {\cal D}{\bf S}_i
Tr e^{-\beta H }
\equiv 
\int {\cal D}{\bf S}_i
e^{-\beta H_{eff} \{ {\bf S} \}}
$$ 
from which it follows that
\begin{eqnarray}
H_{eff} \{ {\bf S}_i \} &=& 
 -{1 \over \beta} logTr e^{-{\beta} H} \cr 
P \{ {\bf S}_i \} &\propto& e^{- H_{eff}\{{\bf S}_i\}} 
\end{eqnarray}
The trace is over the fermion degrees of freedom.
In our case
\begin{equation}
H_{eff} = -{1 \over \beta} logTr e^{-{\beta} H_{el}} + 
J_S\sum_{\langle ij \rangle} {\bf S}_i.{\bf S}_j
\end{equation}

The principal 
difficulty in a  simulation, and quite generally in spin-fermion
problems, is in evaluating the first term on the r.h.s above for an
{\it arbitrary spin
configuration}. 
This is the origin of the $N^3$ factor in the exact MC.
Our key proposal, whose analytic and numerical justification we 
provide later, is 
\begin{equation}
 -{1 \over \beta} logTr e^{-{\beta} H_{el}} \approx
-\sum_{\langle ij \rangle} D_{ij} f_{ij}  
\end{equation}
where $D_{ij}$ is an effective `exchange constant' to be determined as
follows.
Define the operator  ${\hat \Gamma}_{ij}
= (e^{i \Phi_{ij}}  \gamma^{\dagger}_i  \gamma_j + h.c) $. This enters
the `hopping' part of the electron Hamiltonian. 
In any 
specified   spin configuration $\{ f, \Phi \}$ we can 
calculate   
the correlation function 
$D_{ij}\{f, \Phi\} =
Z_{el}^{-1} Tr {\hat \Gamma}_{ij} e^{-\beta H_{el}}$,
where $Z_{el}$ is the electronic partition function in the
specified background. 
The exchange that finally enters $H_{eff}$ is the
average of $D_{ij}\{f, \Phi\}$ 
over the assumed equilibrium distribution, {\it i.e}:
$D_{ij} = \int {\cal D}f {\cal D} \Phi  P \{ f, \Phi  \}
D_{ij} \{ f, \Phi \}
$
where we denote a spin configuration interchangeably by 
 $\{ f, \Phi\}$ or $\{ {\bf S} \}$.
Qualitatively, the `effective exchange' is determined as the
thermal average of a fermion correlator over the assumed
equilibrium distribution.
Let us bring together the equations for ready reference.
\begin{eqnarray}
H_{el}~~~~
&=& -t\sum_{\langle ij \rangle} f_{ij} {\hat \Gamma}_{ij}
+ \sum_i (\epsilon_i - \mu) n_i~~ \cr
\cr
{\hat \Gamma}_{ij}~~~ &=&~~
(e^{i \Phi_{ij}}  \gamma^{\dagger}_i  \gamma_j + h.c) \cr
\cr
f_{ij}~~ &=& ~ \sqrt{( 1 + {\bf S}_i.{\bf S}_j)/2 } \cr
\cr
H_{eff}\{ {\bf S} \}
&=& -{1 \over \beta} logTr e^{-{\beta} H_{el}} + 
J_S\sum_{\langle ij \rangle} {\bf S}_i.{\bf S}_j \cr
\cr
&\approx & -\sum_{\langle ij \rangle } D_{ij} f_{ij} + 
J_S \sum_{\langle ij \rangle} {\bf S}_i.{\bf S}_j \cr
\cr
D_{ij}~~~ &=&~~ \langle \langle {\hat \Gamma}_{ij} \rangle \rangle_{H_{eff}}
\end{eqnarray}

The ED$-$MC approach `solves' for physical properties by 
using the first four  equations above: equilibriating the
spin system by using $H_{eff}$, which itself involves a 
solution of the 
Schrodinger equation for the electrons.

Our method approximates the `exact' $H_{eff}$ by the form
specified in the fifth  equation and {\it computes an  exchange},
rather than equilibrium configurations themselves, by fermion
diagonalisation. The sixth equation indicates  how the
`loop' is closed. We will refer to this method as ``Self Consistent 
Renormalisation'' (SCR) \cite{scr-name},  or the $H_{eff}$ scheme.

The  nonlinear integral equation for the $D_{ij}$ is 
solved to construct the `classical spin model' for a set of
electronic parameters, disorder realisation, and temperature.
Although the assumption about $H_{eff}$ seems `obvious', and 
in fact something similar, but simpler,
had been explored early on by Kubo and
Ohata \cite{kub-ohata}, 
and recently by Calderon and Brey \cite{ed-mc-cal-br}, 
the power of the
method becomes apparent in disordered systems or
in the presence of competing interactions.
In these cases  the 
solutions $D_{ij}$ can be spatially strongly inhomogeneous,
and dramatically temperature dependent. The properties of 
such systems are far from obvious.

The equilibrium  thermal 
average of any  fermion operator, or
correlation function,
${\hat O}$, can
now be computed  
using the self-consistent distribution as:
\begin{equation}
\langle \langle {\hat O} \rangle \rangle
= \int {\cal D} \{{\bf S}\} P \{{\bf S}\} O({\bf S})
\end{equation}
The average 
$O({\bf S})$  is computed on a  
spin configuration  $\{ {\bf S}_i \}$, 
with the  configurations themselves picked according to the
effective Boltzmann weight $\propto e^{-\beta H_{eff}}$.

We have not written the equation for $\mu$. Since we would
typically want to work at fixed density rather than fixed 
chemical potential, we employ the procedure 
above to calculate $n$ and iterate $\mu$ till the `target'
density is obtained. In actual implementation, discussed later,
the $\mu $ `loop' and the $D_{ij}$ `loop' run simultaneously.
We next discuss the analytic underpinning of our method before
moving to numerical results.

\subsection{Analytic limits}

The central problem in DE models is construction 
of an energy functional for arbitrary spin 
configurations  $\{ f, \Phi \}$.
This information is contained in the fermion free energy,
$-T log Tr e^{-\beta H_{el}}$ as we have seen. 
We study two limits below, where the leading effects are
well captured by our effective Hamiltonian.

\subsubsection{ Low temperature}

If we ignore disorder and AF coupling, for simplicity,
and if the free energy of the fermions can be approximated
by the internal energy, then 
$D_{ij} \{f, \Phi\}$ contains the necessary information about the
energy of {\it any spin configuration}: ${\cal E}\{f, \Phi\}
\equiv H_{eff}\{f , \Phi\} = \sum_{ij} D_{ij} \{f , \Phi\} f_{ij}$.
The  {\it configuration dependent}
correlation function, however,
is hard to calculate, since it 
requires 
a solution of the Schrodinger equation for
each spin configuration. 

At low temperature, as the spins gradually randomise, 
the system explores configurations $\{ f, \Phi \}$ near 
the ground state in the  energy landscape. The relevant 
$D_{ij} \{ f, \Phi \}  \sim D^0_{ij}  + \delta D_{ij} \{f , \Phi\}$, 
where $D^0$ is the `exchange' computed on the ground state, and
$\delta D$ is the variation.
At low $T$,
 such that the relevant $\delta D \ll D^0$, 
we can neglect the variation, $\delta D$, between configurations, 
and the `effective Hamiltonian' assumes the form:
$$
lim_{T \rightarrow 0}
~H_{eff} \sim -\sum_{\langle ij \rangle} 
D^0_{ij} f_{ij} =  
-\sum_{\langle ij \rangle} D^0_{ij}
~\sqrt{ (~1 +  {\bf S}_i.{\bf S}_j  )/  2} 
$$
As we will see in the simulations this approximation is 
remarkably good in the simple DE model almost
upto $T_c/2$. 
At higher $T$ the `renormalisation' of $D$ becomes
important.

\subsubsection{High temperature}

For $T_c/T \ll 1$, 
 cumulant expansion yields an
asymptotically exact
effective Hamiltonian: 
$$
H_{eff}~ \sim~ {lim}_{~\beta t \rightarrow 0 }
 -{ 1 \over \beta} 
ln~ Tr ( 1 + \beta H + {1 \over 2} {\beta^2 H^2 } + .. ) 
$$
The leading contribution from this is:
$$
H_{eff}^{high~T} \sim -n (1 - n) \beta t^2 
\sum_{\langle ij \rangle} f^2_{ij}
$$
This apparently has a structure different from that of our $H_{eff}$, and
additionally an `effective coupling' falling off as 
$1/T$. In fact our coupling $D$ has the same form, as can be
checked by evaluating $\langle \langle {\hat \Gamma}_{ij}
\rangle \rangle $ in a high temperature expansion. This
quantity also depends on $n(1-n)$, to allow hopping, and
falls off as $1/T$ since it is non local. The {\it self-consistent}
calculation of the effective exchange, now based on the
high temperature phase rather than the ground state, ensures
that the leading contribution to the energy is well captured.
The physical consequence of the $1/T$ effective exchange is
that the susceptibility of the DE model does not
have the Curie-Weiss form that one expects for Heisenberg like
models \cite{dex-chi-ht}.

The next order in series expansion will generate terms of 
the form:
$$ \sum_{ijkl} 
f_{ij}f_{jk}f_{kl}f_{li}
e^{i (\phi_{ij} + \phi_{jk} + ..)},$$ summed over the minimal plaquette.
Higher powers in $\beta t$ involve longer range excursion of the
fermions, but the limited data available from exact simulations
suggests that the critical properties of  double 
exchange are similar to that of short range spin models. 

Although the procedure above can be extended to extract 
an `exact' effective Hamiltonian to high 
order in $\beta t$, we know of no 
such attempt. 
The only series expansion results available are on the $S=1/2$
model,  directly calculating thermodynamic properties \cite{dex-ht-expn}.

\subsection{Monte Carlo implementation}

Since the ground state of the system is often not known it is
usual to start from high temperature and follow the sequence
below in generating the effective Hamiltonian and studying
equilibrium properties.

$(i)$~We start at high temperature,  $T \gg T_c$, assuming 
some $D_{ij}^n(T)$, where 
$n$ is the iteration index. 
and `equilibriate' the system with this assumed effective Hamiltonian
(not yet self-consistent),
$(ii)$~We compute the  average 
$ \langle \langle 
 e^{i \Phi_{ij}} \gamma^{\dagger}_i \gamma_j + h.c 
\rangle \rangle $ over these  (pseudo) 
equilibrium configurations.
This generates
$D_{ij}^{n+1}(T)$.
$(iii)$~Compare the generated exchange with the assumed exchange at each
bond. Accept if within tolerance. If converged, then $D_{ij}$ represents
the correct `exchange' at that temperature. 
Else, replace $D^n_{ij}$ by $D^{n+1}_{ij}$. 
$(iv)$~At each temperature  and iteration,  
adjust $\mu$ as necessary to keep $n$ constant.

At convergence 
fermion properties can be calculated and averaged over equilibrium
MC configurations of the spin model.
For a disordered system $(\Delta \neq 0)$, 
the thermal cycle
above has to be repeated for each realisation of disorder. In
the clean problem, translation invariance  forces the 
exchange to be uniform at all bonds, while $\Delta \neq 0$ 
generates a  
bond disordered spin model.

The computational effort needed in the ED$-$MC approach is
$\propto N_{MC} \times  N^4$, at each temperature, where  $N_{MC}$ 
is the number of MC sweeps $(10^3-10^4)$, 
and $N$ the size of the system
(actually the Hilbert space dimension).
As we have mentioned before, current resources allow $N_{max}
\sim 100$. Within our $H_{eff}$ scheme the MC configurations
are generated using a short range spin model, with cost ${\cal O}(N)$.
The actual cost is in determining the exchange: this is $\propto
N_{iter} \times N_{av} \times N^3$, 
where $N_{iter}$ is the number of iterations
needed to get a converged solution, with $\sim 10 \%$ accuracy 
per bond, and $N_{av}$ is the averaging needed {\it per iteration}
for generating a reasonable `equilibrium average'.
Typically $N_{iter} \sim 4$ and $N_{av} \sim 50$. 

We can roughly compare the computational cost 
of ED$-$MC with the $H_{eff}$ scheme.
For ED$-$MC, the time required is, $\tau_N \sim
N_{MC} \times N^4$ at a given temperature. For the
$H_{eff}$ scheme, $\tau_{N} \sim 
N_{iter} \times N_{av} \times N^3$.
Putting in the numbers, 
if resources allow $N \sim 100$ for the ED$-$MC approach, the
same resource will allow $N \sim 1000$ within the $H_{eff}$ 
scheme.
In terms of computation time,
$H_{eff}$ is no more expensive than standard
`disorder average' in electronic systems.

\subsection{Physical properties at equilibrium}

The major physical properties we compute at equilibrium are
optical conductivity and d.c resistivity, the density of
states (DOS), and the magnetic structure factor.

$(i)$~We estimate the d.c conductivity, $\sigma_{dc}$,
by using the Kubo-Greenwood
expression \cite{mahan} for the optical conductivity.
In a disordered non interacting system we have:
\begin{equation}
\sigma ( \omega)
=  {A \over N}
\sum_{\alpha, \beta} (n_{\alpha} - n_{\beta})
{ {\vert f_{\alpha \beta} \vert^2} \over {\epsilon_{\beta}
- \epsilon_{\alpha}}}
\delta(\omega - (\epsilon_{\beta} - \epsilon_{\alpha}))
\end{equation}
The constant $A=
 (\pi e^2)/  {\hbar a_0 }$.
The matrix element $f_{\alpha \beta} = \langle \psi_{\alpha}
\vert j_x \vert \psi_{\beta} \rangle$ and we use  the current operator
$j_x = i  a_0  \sum_{i, \sigma} (c^{\dagger}_{{i + x a_0},\sigma}
c_{i, \sigma} - h.c)$. The $\psi_{\alpha}$ etc
 are single particle eigenstates,
for a given equilibrium configuration, and $\epsilon_{\alpha},
\epsilon_{\beta}$
are the
corresponding eigenvalues. The
$n_{\alpha}= \theta(\mu
- \epsilon_{\alpha})$, etc, are occupation factors.
 
The conductivity above is prior to thermal or
disorder averaging.
Our simulations are in a  square or cube geometry
with periodic boundary condition.
Given the
finite size, the $\delta$ function constraint in $\sigma(\omega)$
cannot be satisfied
for arbitrary $\omega$. We use the following strategy:
$(i)$~calculate $\sigma_{int}(\omega)
= \int_0^{\omega'} \sigma(\omega') d \omega'$,
at three equispaced
low frequency points, $\omega_1, \omega_2,  \omega_3$,
by summing over the delta functions in the appropriate range..
$(ii)$~thermally average the $\sigma_{int}(\omega)$ over the
equilibrium configurations,
$(iii)$~invert: calculate a numerical derivative via
three point interpolation, implementing
${\bar \sigma}(\omega) = d {\bar \sigma}_{int}(\omega)/d\omega$.
The `bar' on $\sigma$ indicates thermal average.
What we call the `d.c. resistivity' is actually the inverse of
a low frequency optical conductivity, computed by the method
above. We systematically check the stability of our results by
repeating the calculation for a sequence of system size (and
reducing $\omega_1, \omega_2, \omega_3$ accordingly).
For $N \sim 1000$, the `d.c' conductivity is actually
computed at $\omega \sim 0.06$.

Our transport calculation method
and some benchmarks will be
discussed in detail elsewhere \cite{sk-pm-transp}.
To convert to `real' units, note that
our conductivity results are in units
of $(\pi e^2)/{\hbar a_0}$.
Since the Mott `minimum' metallic conductivity, in three
dimension, is $\sim (0.03 e^2/\hbar a_0)$, $\sigma = 1$ on our scale
roughly corresponds to $10^2 \sigma_{Mott}$.
The full $\sigma(\omega)$ is computed by computing $\sigma_{int}
(\omega)$ defined above, thermal average, and inversion.

$(ii)$~Each equilibrium magnetic configuration leads to a `DOS' of the form
$\sum_{\alpha} \delta(\omega - \epsilon_{\alpha})$, where $\epsilon_{\alpha}$
are the single particle eigenvalues in that background. 
The
thermally averaged DOS that we show involves a 
Lorentzian 
broadening of each  $\delta$ function, as indicated below. 
\begin{equation}
N(\omega) \approx { 1 \over {N_{eq}}}
\sum_{eq}
\sum_{\alpha} { (\Gamma/\pi) \over { 
(\omega - \epsilon_{\alpha})^2 + \Gamma^2 }}
\end{equation}
The sum runs over the eigenvalues obtained in
any spin configuration, and summed over equilibrium configurations.
We use $\Gamma \sim 0.1$ in our results, although much
smaller $\Gamma$ would still give a smooth spectra at high
$T$.

$(iii)$~The magnetic structure factor is calculated as
\begin{equation}
S({\bf Q}) = 
{ 1 \over {N_{eq} N^2}}
\sum_{eq} \sum_{ij} \langle {\bf S}_i.{\bf S}_j \rangle
e^{i {\bf Q}.({\bf r}_i -{\bf r}_j)}
\end{equation}
where $i,j$ run over the entire lattice, and the outer average is over
equilibrium configurations.

\section{Results}

In this section we provide a comprehensive comparison of results
based on the `exact' scheme (ED$-$MC) and our effective Hamiltonian approach,
for the `clean' DE model, and  
and extend the study to large sizes using the $H_{eff}$ scheme. 
Most of our results are on three dimensional systems, where the simulations are
more difficult and the results physically more relevant, and we show
only limited data in two dimensions.
The model is translation invariant, there 
are no competing interactions, and the low temperature phase is a 
ferromagnet.

\subsection{Magnetism and thermodynamics}

We begin with a comparison of the magnetisation, $m(T)$, obtained via
ED-MC and SCR  
on $8 \times 8$ lattices in 2d, and $4^3$ systems in 3d,
with periodic
boundary condition 
in all directions. Fig.1 compares 
the $m(T)$ obtained
via the two schemes at three electron densities.

Note at the outset that both the DE  model and our
$H_{eff}$ are $O(3)$ symmetric and are {\it not } expected to
have long range order at finite $T$ in 2d (in an infinite
system). 
However, as has been demonstrated in the case of the two dimensional
classical Heisenberg model \cite{takahashi}, $O(3)$ models
have {\it exponentially large} correlation length at low 
temperature in 2d. 
For a nearest neighbour classical
Heisenberg  model
with $\vert {\bf S}_i \vert =1$, and exchange $J$,
the low $T$ 
correlation length $\xi (T) \sim 0.02 e^{2 \pi J/T}$.
So, for $T \ll J$ even large finite lattices would look `fully
polarised' and one would need to access exponentially large sizes
to see the destruction of long range order. 

This allows us to
define a (weakly size dependent) `characteristic temperature'
$T_{ch}(n)$ for the 2d DE model which marks the crossover from
paramagnetic to a nominally `ordered' phase. The true ordering 
temperature of strongly anisotropic DE systems,
{\it e.g}, 
the layered manganites, which the planar model mimics,
would be determined by the
interplane coupling, but the in plane transport would
be dictated mainly by the 2d fluctuations,~as~here.

The difference between ED-MC and SCR results in 2d,  Fig.1.(a),
 is most prominent at the highest
density, $n=0.41$, where the $T_{ch}$ inferred from these small size
calculations differ by $15 \%-20 \%$. 
At lower density the difference is
still visible but much smaller. We  have indicated the $T_{ch}$ 
scales inferred from the two schemes in the inset in Fig.1.(a)
The difference between
the two schemes is usually largest in clean high density systems,
as we will see also in the three dimensional case.
However, over the entire density range, the maximum deviation
is $\sim 20 \%$.

\begin{figure}
\vspace{1.0cm}
\centerline{
\psfig{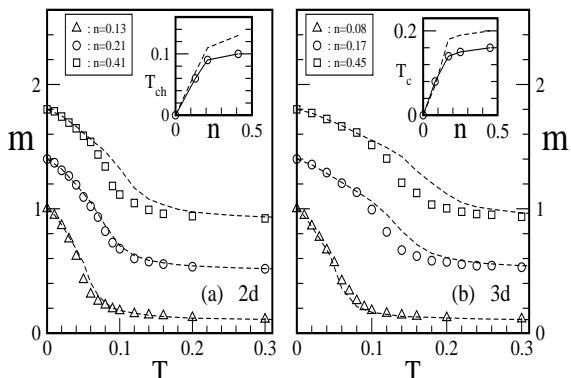}}
\caption{ Magnetism in 2d and 3d:
open symbols are for ED$-$MC, the dotted lines indicate the SCR
results, $(a).$~2d, $(b).$~3d. 
The insets show the $T_c$ obtained via ED-MC (symbols) -vs-
SCR results (dotted lines).}
\end{figure}
\begin{figure}
\centerline{
\psfig{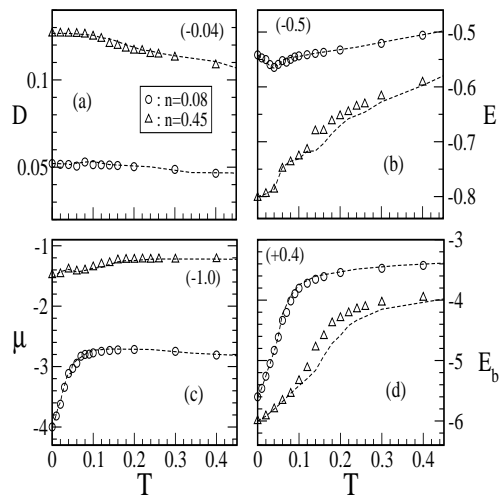}}
\vspace{.3cm}
\caption{ Comparing thermodynamic indicators between ED-MC and the
$H_{eff}$ scheme in 3d: 
$(a).$~effective exchange, $(b).$~internal
energy, $(c).$~chemical potential and $(d).$~band edge. 
Displayed value is actual value $+$ shift. System size $4 \times 6 \times 4$.
Open symbols: ED-MC, dotted lines: SCR.}
\end{figure}

Notice that at all $n$, the low temperature $m(T)$ obtained via
$H_{eff}$ corresponds almost exactly with results based on
ED$-$MC. This works upto $\sim T_{ch}/2$.
The high temperature result within the two schemes 
is also in close 
correpondence but that is better illustrated in the
thermodynamic data, Fig.2,
which we will discuss later.

Fig.1.(b)  shows the results on magnetisation in the three dimensional
problem at three densities, comparing results based on ED$-$MC
and $H_{eff}$.
As in two dimension the difference in the estimated
$T_c$ is greatest near the band center, being $\sim 15 \% - 20 \%$,
the correspondence improving as we move to $n \lesssim 0.2$. As before,
the exact and approximate $m(T)$ match at low $T$ for all densities.
 
Fig.2 which shows the thermodynamic indicators in the
3d case  reveals that $D_{ij}$  itself 
is virtually indistinguishable
in the two schemes.
The correlation $ D_{ij} =
\langle \langle \hat 
\Gamma_{ij} \rangle \rangle $
can be evaluated as an equilibrium average in an exact simulation 
also, although there 
it does not feed back into the calculation. 
The match between the $D$'s computed in two different schemes,
and across the density range, suggests that the difference
in $m(T)$ seen near half-filling is not due to different numerical
values of $D$, but the assumed {\it form } of $H_{eff}$.
We either need a more sophisticated definition of the finite
temperature $D$, or a different form of $H_{eff}$ to bring the
high density results of $H_{eff}$ in closer correspondence
with ED$-$MC.
Notice that the $D$'s are only weakly temperature dependent
and the $m(T)$ at low temperature could have been obtained by setting
$D(T) = D(0)$.
In fact over the temperature 
range $0-T_{c}$ the qualitative physics can
be accessed without the thermal `renormalisation' 
of the exchange. However, for $T \gg T_{c}$ the renormalisation
is important, as suggested earlier by the high temperature
expansion.

The results on all
thermodynamic indicators,  $D(T)$, $E(T)$, $\mu(T)$ and
$E_b(T)$, Fig.2,
show the close correspondence between results
of the exact and approximate scheme. 
The $D$'s are almost temperature independent in
the range $0-T_c$ and hardly distinguishable
between ED$-$MC and $H_{eff}$, suggesting that effects beyond
our effective Hamiltonian
$-D \sum f_{ij}$ is needed to accurately describe the magnetic
transition at the band center.
The overall behaviour is similar in 2d as well
so we are not
presenting the 2d data.

We  extend the $H_{eff}$ scheme to large system
size, and study the magnetism in $32^2$ and $10^3$ lattices.
Fig.3 shows the results on $m(T)$, and
the inset shows the $T_c$ inferred from these simulations.
The maximum $T_c$, occuring at band center is $\sim 0.2t$ which,
with $t \sim (100 - 150)$meV, will be in the range $200-300$K.
These numbers are typical of high electron density Hunds
coupled systems, and are in the right ballpark when compared
to the manganites \cite{tc-comp}.

\begin{figure}
\vspace{1.5cm}
\centerline{
\psfig{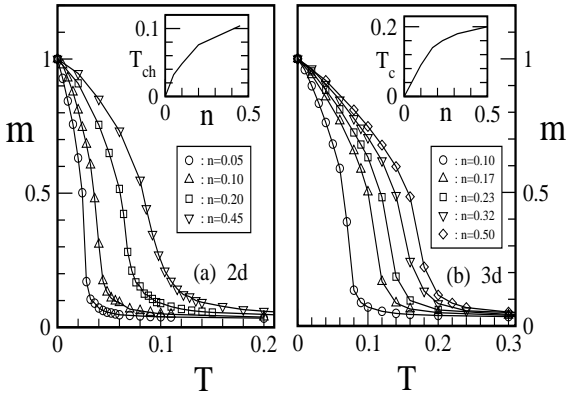}}
\vspace{.3cm}
\caption{ Magnetisation based on $H_{eff}$ in $(a).$2d with
~$30 \times 30$
 and $(b).$~3d with $ 10 \times 10 \times10 $ systems. Insets show the
 characteristic temperature scales inferred from $m(T)$
}
\end{figure}
\begin{figure}
\centerline{
\psfig{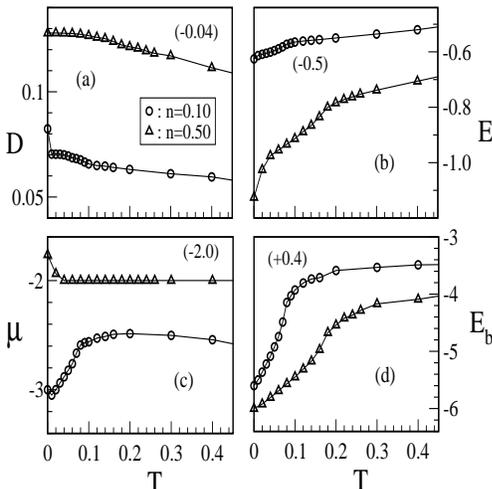}}
\caption{ Thermodynamic properties in the 
3d case computed with $H_{eff}$, system size $N= 10 \times 10 \times 10$.
}
\end{figure}
\begin{figure}
\vspace{1.5cm}
\centerline{
\psfig{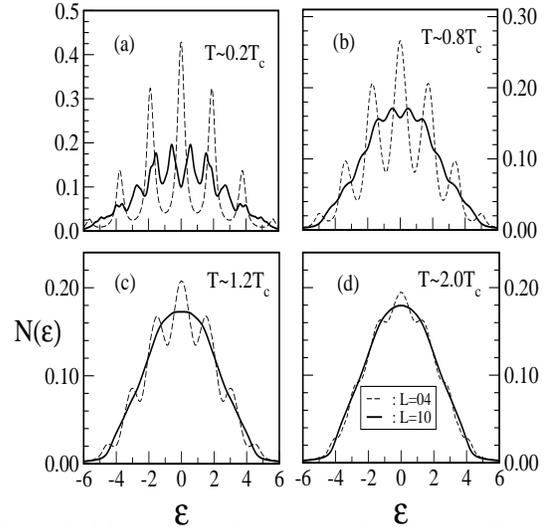}}
\caption{ DOS in three dimension. Results on $H_{eff}$ 
with $4 \times 6 \times 4$
and $10 \times 10 \times 10$ geometry, $n=0.3$. 
}
\end{figure}

Fig.4 shows the thermodynamic indicators computed within the
$H_{eff}$ scheme on $10^3$ in 3d. 
The strong temperature  dependence in $\mu$ and $E_b$, seen also
at small sizes,  arise 
from the `band narrowing' effect of spin disorder which reduces
the mean hopping amplitude with increasing temperature.

\subsection{Density of states}

Fig.5 shows the density of states (DOS) computed at
$n=0.3$, four temperatures, and for a small, $4^3$, 
and a large, $10^3$, system.
The mean level spacing at high temperature (where
the spins are completely disordered)  is
$\sim 12/L^3$ 
which is $\sim 0.01$ at $L=10$ and $\sim 0.18$
at $L = 4$. 
For $T \rightarrow 0$, the polarised ferromagnetic state
leads to
large degeneracy and the level spacings could
be more than $10$ times larger than 
the high temperature value.
We have broadened all $\delta$ functions by $\Gamma = 0.1$,
so that the high temperature $L=4$ spectra looks reasonable.
With this broadening the $L=10$ data looks
reasonable even below $T_c$.

This  comparison  highlights
the unreliability of small size data in inferring spectral
features over most of the interesting temperature range. 
Small sizes can often provide reasonable results  on
energetics, but on spectral features and, more importantly,
on low frequency transport, they are completely unreliable.

\subsection{Optical properties}
Fig.6 shows the optical conductivity, $\sigma(\omega)$.
The optical conductivity is a vital probe of charge dynamics
in the system. Our data in the main panel, Fig.6, is for
a $8 \times 8 \times 8$ geometry. At the lowest temperature
there is an artificial `hump' in $\sigma(\omega)$ which 
we think arises because the polarised 
three dimensional system has
large degeneracy, and 
finite size effects
are  stronger than in two dimension.
Nevertheless, there are some notable features in $\sigma(\omega)$,
$(i)$~the conductivity is Drude like,
$(ii)$~there  is
rapid reduction in
low frequency spectral weight with increasing temperature, 
with some transfer to high frequency, 
$(iii)$~the weight in
$\sigma(\omega)$ is not conserved with increasing temperature,
the loss is related to the suppression of kinetic energy with
increasing spin disorder.

\begin{figure}
\vspace{1.0cm}

\centerline{
\psfig{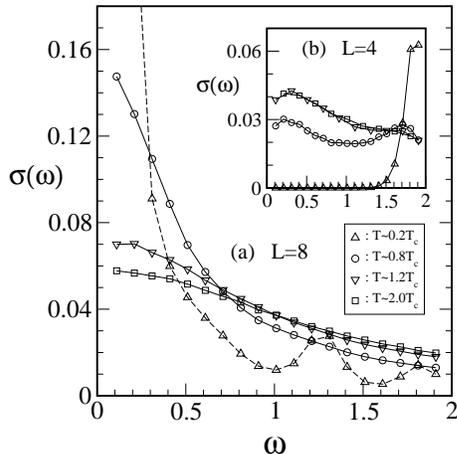}}
\caption{ Optical conductivity based on $H_{eff}$ in three 
dimension. System size $4 \times 4 \times 4$ (inset) and $8 \times 8
\times 8$ (main panel), density $n=0.3$. Symbols in the inset are same as
in the main panel.}
\end{figure}
\begin{figure}
\vspace{.5cm}

\centerline{
\psfig{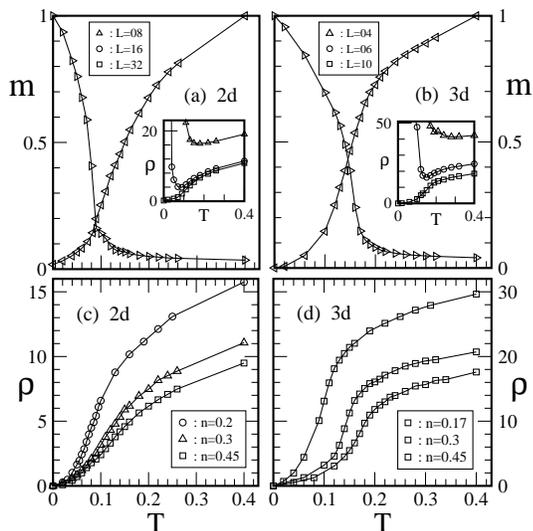}}
\caption{$(a)-(b).$~Magnetisation and normalised 
resistivity at 
$n=0.3$  in $(a).$~2d and
 $(b).$~3d. Insets to $(a).$ and
 $(b).$ show the size dependence in the resistivity (see text). 
$(c).-(d).$~Density dependence of $\rho(T)$, in 2d and
3d respectively, with system sizes  
~$ 32 \times 32 $ and $(d)$ ~$ 10 \times 10 \times 10$.}
\end{figure}
\begin{figure}
\centerline{
\psfig{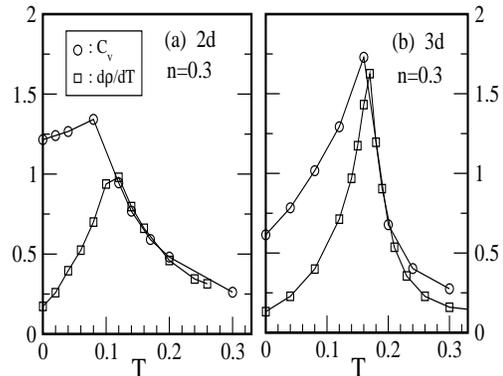}}
\caption{Specific heat and $d \rho/dT$, at $n=0.3$ in $(a).$~2d, and 
$(b).$~3d. System sizes used are same as in panels 7.$(c)-(d).$ }
\end{figure}

\subsection{Resistivity}

Finally, we look at the resistivity, which,
surprisingly, has seen little discussion. Fig.7 shows
the correlation between the ferromagnet to paramagnet 
transition and the rise
in $\rho(T)$. We have normalised $\rho(T)$ 
in Fig.7.(a)-(b) by the value at
$T=0.4$. The `absolute' resistivity is shown in Fig.7.(c)-(d).
Unlike mean field treatments which treat the paramagnetic
phase as completely `uncorrelated' and would yield a `flat'
resistivity for $T > T_{ch}$ (or $T_c$),
 there is a significant increase
in $\rho(T)$ with rising temperature in the `paramagnetic'
 phase
as the short range spin correlation is gradually lost and
the system heads towards the fully spin disordered phase. 
The general rise in $\rho(T)$ in the paramagnetic phase 
happens in both 2d and 3d, but 
surprisingly in
2d most
of the rise seems to occur after the drop in $m(T)$, rather than
across $T_{c}$ as one sees in three dimension.

For a check on the reliability of the computed $\rho(T)$
the inset in Fig.7.(a) shows the `resistivity' computed on
$L \times L$ geometry for $L = 8, 16,  32$ across
the full temperature range. The $L=8$ result has the 
same problem that we discussed in the context of 
$\sigma(\omega)$. The system essentially behaves as an 
`insulator' at low $T$ due to the finite size gap. The
$L=16$ data has similar upturn, but at a lower temperature.
The data at 
$L=24$ (not shown) and $L= 32$ are stable down to $T \sim 0.02$ 
and 
almost coincide, 
suggesting that except at very low temperature, results on
these sizes
are representative of bulk transport.

The resistivity in the 3d case differs from 
2d in
that the major rise in $\rho(T)$ occurs around $T_c$ in the
3d case, while it occurs {\it beyond} $T_{ch}$ in the 2d case. Fig.7.(b) 
shows $m(T)$ correlated with the normalised $\rho(T)$, and the
rise is reminiscent of the Fisher-Langer result \cite{fish-lang}
 in  weak
coupling electron-spin systems.  
The inset in Fig.7.(b) shows the stability of the transport result in
3d for
$L \gtrsim 8$, and the unreliability for $L \sim 4$. 

Fig.7.(c)-(d), shows the absolute resistivity for a few densities. 
The `high temperature' 3d resistivity, at $T \sim 3T_c$ is approximately
$15-25$, in the density range shown, which in real units would
be $\sim (1-2) m\Omega$cm, roughly the high $T$ resistivity
of La$_{1-x}$Sr$_x$MnO$_3$ for $x \gtrsim 0.4$.

Fig.8. shows the correlation between $d\rho/dT$
and the specific heat in 2d and 3d. Above $T_c$ and in 3d,
panel (b),
$d\rho/dT$ seems to match $C_V$ very
well, as expected from the perturbative results of Fisher and Langer 
\cite{fish-lang}. In 2d however the correspondence is poor, probably
due to incipient localisation effects in the resistivity.
For $T \le T_c$, even in 3d, the behaviours of $C_V$ and
$d\rho/dT$ are different because the rise in $m(T)$ affects
the scattering rate, as is already known \cite{fish-lang}. 

The  validity of
the `weak coupling' results of Fisher-Langer, originally illustrated
for a Heisenberg model, in this  
`strong coupling' spin-fermion system may seem 
surprising. There are two reasons why the correspondence holds here:
$(i)$~the resistivity in the DE model arises from 
spin disorder induced weak fluctuations in the hopping amplitude, and
is in the perturbative regime, 
and $(ii)$~our magnetic model, $H_{eff}$,
is effectively short range, and the critical properties of 
spin fluctuations are the same as in the Heisenberg model.

\section{Conclusion}

In this paper we proposed a 
new Monte Carlo technique
that allows access to large system sizes but retains
the correlated nature of spin fluctuations in the
double exchange 
model. Combining this MC technique with a transport calculation
based on the exact Kubo formula we 
presented a comprehensive solution of the  model, including 
magnetism, thermodynamics, spectral features, transport, and 
optics. 

This paper benchmarked the scheme 
for the clean double exchange model, where the complicated
consistency and thermal renormalisation involved in the scheme
are not crucial for a qualitative understanding. 
However, when we move to disordered systems \cite{sk-pm-dde},
 or non ferromagnetic
ground states, or the regime of multiphase coexistence \cite{sk-pm-nano1},
 the full
power of a `bond disordered' effective Hamiltonian, with non trivial
spatial correlation between the bonds, becomes apparent. 

For the clean ferromagnetic case one may try to improve
the self-consistency scheme to obtain better correspondence
\cite{ed-mc-cal-br,solov}
with ED$-$MC results. 
However, given the complexity
of the current scheme, and the range of possibilities that it
offers, we think it is more important to exploit the present
scheme to resolve the outstanding {\it qualitative issues} first.
Finally, although the entire scheme is
presently implemented numerically,
it would be useful to make analytic approximations 
within this framework
to create greater qualitative understanding.

\vspace{.2cm}

\noindent
We acknowledge use of the Beowulf cluster at HRI.


\begin{thebibliography}{99}
\bibitem{scr-name} Our method should not be confused with the 
`self-consistent renormalisation' scheme developed by T. Moriya and
coworkers in the context of spin fluctuations in $d$ electron systems.
\bibitem{tc-comp}
Comparing  with results of other recent MC techniques,
at $n=0.5$ the $L\rightarrow \infty$ extrapolation of both 
hybrid MC\cite{nucl-phys}
and polynomial expansion MC \cite{furu-new} give
$T_c \sim 0.14t$. We do not have finite size scaling data but
can guess that our $L \rightarrow \infty$ $T_c$ will be $\sim 0.18t$.
\end{thebibliography}
\end{document}